\documentclass[12pt,preprint]{aastex}
\usepackage{graphicx}
\usepackage{natbib}
\newcommand{\intunr}{\mbox{$I_{\rm CXB}^{\rm unres}(20-50\;{\rm keV})$}}
\newcommand{\inttot}{\mbox{$I_{\rm CXB}^{\rm tot}(20-50\;{\rm keV})$}}

\shorttitle{The cosmic X--ray background with BSAX/PDS}
\shortauthors{Frontera et al.}
\slugcomment{{\em Astrophys.\ J. Part 1: accepted}}

\begin{document}

\title{The cosmic X--ray background and the population of the most heavily obscured AGNs}

\author{Filippo Frontera\altaffilmark{1,2}, 
 Mauro Orlandini\altaffilmark{1}, Raffaella Landi\altaffilmark{1},
 Andrea Comastri\altaffilmark{3}, 
 Fabrizio Fiore\altaffilmark{4},
 Giancarlo Setti\altaffilmark{5,6},
 Lorenzo Amati\altaffilmark{1},  
 Enrico Costa\altaffilmark{7}, 
 Nicola Masetti\altaffilmark{1},
 and Eliana Palazzi\altaffilmark{1}}

\altaffiltext{1}{INAF/Istituto di Astrofisica Spaziale e Fisica Cosmica,
  Bologna, Via Gobetti 101, 40129 Bologna, Italy (email: frontera@fe.infn.it)}
\altaffiltext{2}{Dipartimento di Fisica, Universit\`a di Ferrara, 
  Via Saragat, 1, 44100 Ferrara, Italy}
\altaffiltext{3}{INAF/Osservatorio Astronomico di Bologna, Via Ranzani, 1, 
  40127 Bologna, Italy}
\altaffiltext{4}{INAF/Osservatorio Astronomico di Roma, 00040 Monte Porzio 
  Catone, Italy}
\altaffiltext{5}{Dipartimento di Astronomia, Universit\`a di Bologna, Via 
  Ranzani, 1, 40127 Bologna, Italy}
\altaffiltext{6}{INAF/Istituto di Radioastronomia di Bologna, Via Gobetti, 
  101, 40129 Bologna, Italy}
\altaffiltext{7}{INAF/Istituto di Astrofisica Spaziale e Fisica Cosmica,
  Roma, Via del Fosso del Cavaliere 100, 00133 Roma, Italy}

\begin{abstract}
We report on an accurate measurement of the CXB in the 15--50 keV range 
performed with the {\em Phoswich Detection System (PDS)} instrument aboard 
the {\em BeppoSAX} satellite. We establish that the most likely CXB 
intensity level at its emission peak  (26--28 keV) is $\approx$40 keV
cm$^{-2}$~s$^{-1}$~sr$^{-1}$, a value consistent with that derived from 
the best available CXB measurement obtained over 25 years ago with the 
first {\em High Energy Astronomical Observatory satellite mission} 
\citep[{\em HEAO--1};][]{Gruber99}, whose intensity, lying well below the
extrapolation of some lower energy measurements performed with focusing telescopes,
was questioned in the recent years. We find that 90\% of the acceptable
solutions of our best fit model to the PDS data give  a 20--50 keV CXB flux 
lower 
than $6.5\times 10^{-8}$~erg~cm$^{-2}$~s$^{-1}$~sr$^{-1}$, which is 12\% higher 
than that quoted by \citet{Gruber99} when we use our best calibration scale. This
scale gives a 20--50 keV flux of the Crab Nebula of $9.22 
\times10^{-9}$ erg~cm$^{-2}$ s$^{-1}$, which is in excellent agreement with the
most recent Crab Nebula measurements and  6\% smaller than that assumed 
by \citet{Gruber99}. In combination 
with the CXB synthesis models we infer that about 25\% of the 
intensity at $\sim 30$~keV arises from extremely obscured, Compton thick AGNs 
(absorbing column density $N_{\rm H}>10^{24}$~cm$^{-2}$), while a much 
larger population would be implied by the highest intensity estimates. 
We also infer a mass density of supermassive BHs of $\sim 
3\times 10^{5}$~M$_{\odot}$~Mpc$^{-3}$. The summed contribution of 
resolved sources \citep{Moretti03} in the 2--10 keV band exceeds our best 
fit CXB intensity extrapolated to lower energies, but it is within 
our upper limit, so that any 
significant contribution to the CXB from sources other than AGNs, such as 
star forming galaxies and diffuse Warm-Hot Intergalactic Medium (WHIM), is 
expected to be mainly confined below a few keV.
\end{abstract}


\section{Introduction}

The cosmic X-ray background (CXB) is contributed mainly by active galactic 
nuclei (AGN) powered by accreting supermassive black holes at the centers 
of large galaxies~\citep{Setti89,Comastri95,Gilli04}. Optically bright 
quasars and Seyfert galaxies dominate at low energies (up to a few keV), 
while obscured AGNs, which outnumber unobscured ones by a factor 3--4 
\citep{Ueda03,Lafranca05}, are responsible for the bulk of the CXB at high 
energies ($>$10 keV). However the CXB intensity level is still a matter of 
debate. After the first pioneer CXB measurements~\citep{Horstman75}, the 
major effort to get a reliable estimate of the spectrum in a broad energy 
band (2--400 keV) was performed in the late 1970's with the A2 and A4 
instruments aboard the first {\em High Energy Astronomical Observatory} ({\em 
HEAO--1\/}). The A2 results (3--45 keV) were first presented by 
\citet[ hereafter M80]{Marshall80}, while the final results obtained with 
the A4 Low Energy Detector (LED, 13--180 keV) were reported by \citet[ hereafter
G99]{Gruber99}, who also presented the conclusive results from both experiments. 
According to these authors the CXB energy spectrum 
$J(E)$ in the 3--60 keV interval is well represented by a 
power-law (PL) with a high energy exponential cutoff ({\sc cutoffpl}), while the 
corresponding $E\,J(E)$ spectrum shows a characteristic bell shape 
with a maximum intensity of 42.6 keV (cm$^2$~s~sr)$^{-1}$ at 29.3 keV.

After {\em HEAO--1} there have been many other CXB measurements  
at low energies ($<$15 keV) with both imaging and non imaging telescopes 
aboard satellite missions, but at high energies ($>$15 keV) no
accurate measurement has been published yet. A major effort has been recently
performed with {\em INTEGRAL} \citep{Churazov06}. In this case the CXB measurement
was obtained by leaving the Earth disk 
(angular size of $\sim 5.4^\circ$ as seen by the satellite) to cross the
much larger field of view (FOV) of the satellite mask telescopes and
fitting the depression in the count rate with a multi-component model, in which
one of the components was the CXB and the others were several, e.g.,
the Earth X-ray albedo, the instrumental background, the celestial sources 
in the FOV. The many assumptions about the contribution, at the epoch of 
the measurement,  of all the components to the depression level and its dependence 
on energy and time, make this measurement somewhat difficult for an unbiased 
estimate of the CXB. Indeed \citet{Churazov06} wish further INTEGRAL 
observations at other epochs "to verify the agreement of observations and 
predictions".
 
In Table~\ref{t:CXB_le} we compare the 
major CXB results.  At low energies (1--15 keV), it is 
apparent a low spread of the {\sc pl} photon index $\Gamma$ and a high 
spread (up to $\sim$40\%) of the CXB intensity, exemplified by the ratio 
$R_{\rm HEAO-1}^{2-10\;{\rm keV}}$ between the measured 2--10 keV 
intensities and that measured with {\em HEAO--1}, with the lowest CXB 
estimates obtained with {\em HEAO--1} A2 (M80) and the 
highest with the focusing telescopes aboard {\em BeppoSAX} 
\citep{Vecchi99}, {\em XMM-Newton} \citep{Lumb02,Deluca04} and {\em 
Chandra} \citep{Hickox06}. The ratio ($= 1.15$) obtained with the collimated 
{\em Proportional Counter Array} aboard 
{\em Rossi--XTE\/} \citep{Revnivtsev03} and that ($= 1.19$) obtained from 
the re-analysis of the {\em HEAO--1} A2 measurement \citep{Revnivtsev05} 
can be lowered to 1.04 and 1.07, respectively, by a more reliable 
calibration of the flux scale. Indeed the adopted 2--10 keV Crab flux 
\citep{Zombeck90} is $\sim$11\% higher than the mean value of all other 
Crab flux estimates. In the 20--50 keV energy band, which is common to the
high energy X--ray experiments, the few available 
measurements show a low spread of the ratio $R_{\rm HEAO-1}^{20-50\;{\rm 
keV}}$ but systematic errors in the CXB intensity estimates cannot be 
excluded.

Driven by these discordant results several authors 
\citep{Ueda03,Deluca04,Comastri04,Worsley05,Ballantyne06,Hopkins06,Worsley06}, 
in their evaluation of the fraction of the CXB that can be resolved into 
individual sources or in their CXB source synthesis models, assume the 
3--60 keV CXB spectrum obtained with {\em HEAO--1} (G99) to be 
corrected in its shape, but underestimated in its intensity due to 
systematic errors in the absolute area calibration and/or instrumental 
background subtraction.
As a result, the {\em HEAO--1} CXB intensity is 
increased upward by a factor up to $\approx$1.3--1.4 over the entire energy band.

In order to establish whether such CXB intensity renormalization is 
justified, and thus to constrain the size of a population of highly 
obscured AGNs and to infer the presence of other source populations and/or 
of a truly diffuse component \citep[WHIM, see][]{Kuntz01b}, we have 
performed an accurate measurement of the total (resolved plus unresolved) 
high energy ($>$15 keV)  CXB intensity by exploiting the pointed 
observations performed with the {\em Phoswich Detection System} (PDS) 
aboard the {\em BeppoSAX} satellite \citep{Boella97b}.

\section{Instrument and calibrations}
\label{s:cal}
A detailed description of the PDS instrument and its in--flight 
performance can be found elsewhere \citep{Frontera97b,Frontera97a}. 
Although the PDS was not designed to perform a measurement of the CXB 
(Field of View, FOV, of only 1\fdg3 FWHM), {\em a posteriori} it was 
realized that such measurement would be possible given the very good 
performance of the instrument: high temporal stability, very low 
background level $B_{\rm 15-300\;keV} = 1.6\times10^{-4}$ counts 
(cm$^2$~s~keV)$^{-1}$, high flux sensitivity and good energy calibration.  
The 15--300 keV limiting sensitivity corresponds to about 1\% of the 
background level with a marginal influence (0.3\%) of systematic errors in 
the background subtraction, also thanks to the continuous monitoring of 
the background with two rocking collimators which alternated with a 
default dwell time of 96~s between the neutral position (ON--source) and 
two default symmetrical positions offset by $\pm 3\fdg5$ 
($\pm$OFF--source).

The ON--axis PDS response function was determined by means of pre--launch 
calibrations combined with Monte Carlo calculations, and it was tested 
during the {\em BeppoSAX} life time (6 yrs) with 7 repeated observations 
of the Crab Nebula (twice in 1997 and 1999, and once in the years 1998, 
2000, and 2001) that were performed simultaneously with the other Narrow Field 
Instruments (NFIs) on board. By assuming a power--law ({\sc pl}) model 
of the form $I(E) = N E^{-\Gamma}$~photons~cm$^{-2}$~s$^{-1}$~keV$^{-1}$ 
it was found that the photon index 
$\Gamma$ obtained with the PDS remained unchanged in all of these 
observations within the statistical uncertainties, with a mean value of 
$2.121\pm 0.001$ and a reduced $\chi^2/{\rm dof} = 25.5$ for 66 degrees of freedom 
(dof) in the 15--200 keV band. A better fit ($\chi^2/{\rm dof} = 17.1$ 
for 64 dof) in this energy band was obtained with a broken power-law ({\sc 
bknpl}) with a mean low energy index $\Gamma_1 = 2.113\pm 0.001$, a break 
energy $E_{\rm b} = 74\pm 2$~keV and a high energy index $\Gamma_2 = 
2.198\pm 0.005$. A {\sc bknpl} spectrum was also the best fit to other 
broad band high energy measurements \citep[see, e.g.,][]{Bartlett94}. The 
still high $\chi^2/{\rm dof}$ obtained with the {\sc bknpl} model can be 
lowered and made compatible with the $\chi^2$ statistics if a systematic 
error of only 1\% in the used PDS response function is assumed.

\begin{figure}
\begin{center}
\includegraphics[width=\textwidth]{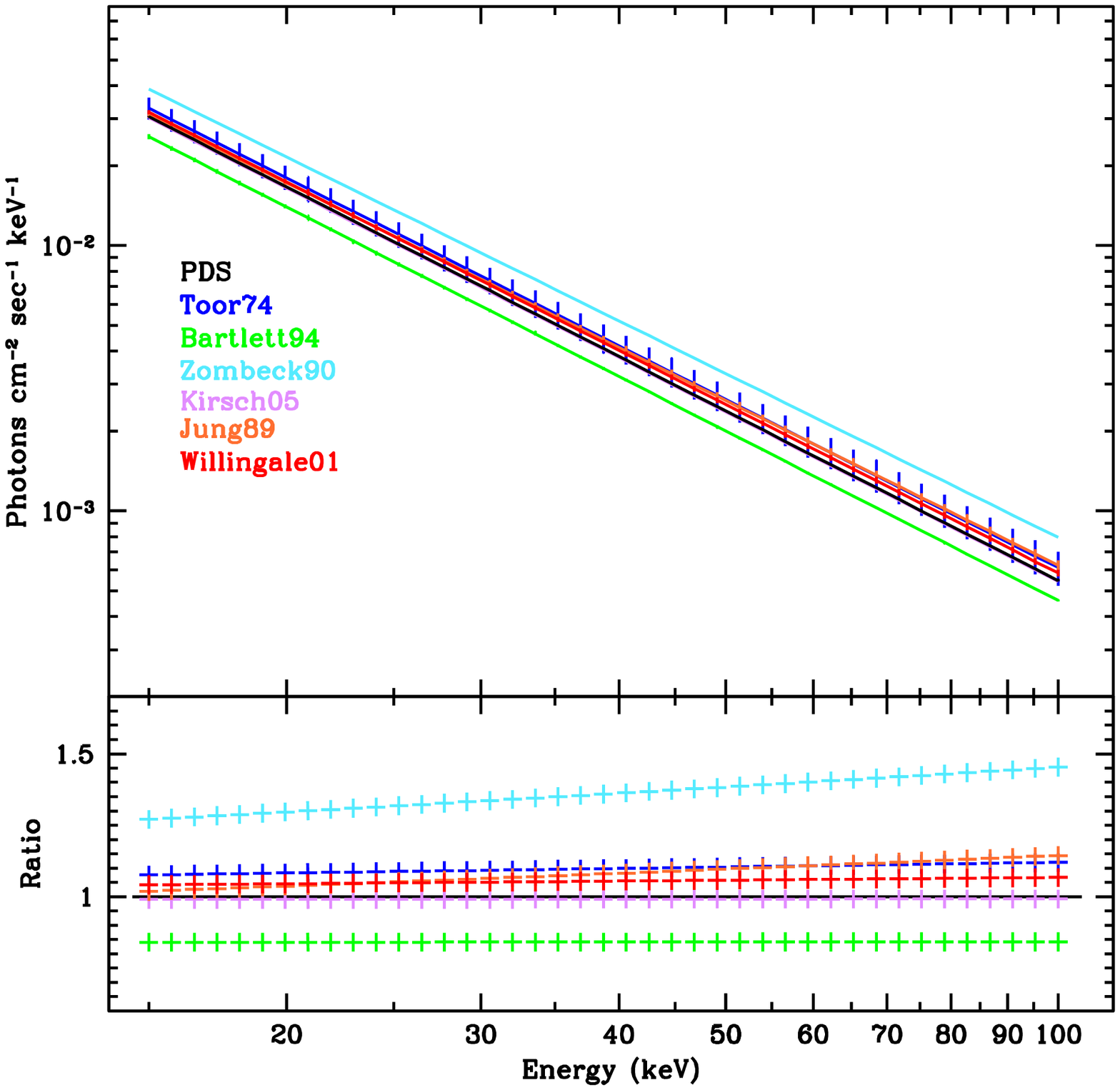}
\end{center}
\caption[]{{\em Upper panel:} {\em BeppoSAX} 15--100 keV Crab spectrum 
(black points) compared with other measurements. {\em Blue line}: review of 
different measurements in 5--70 keV \citep{Toor74}; {\em Green line}: GRIS 
balloon experiment in 20--1000 keV \citep{Bartlett94}; {\em Cyan line}: 
\citet{Zombeck90}; {\em Purple line}: collection of different measurements 
in 2--50 keV \citep{Kirsch05}; {\em Orange line}: 15--180 keV {\em 
HEAO--1} A4/LED measurement \citep{Jung89}; {\em Red line}: XMM-Newton
\citep{Willingale01}. When not visible, error bars 
are smaller than the line thickness. {\em Bottom panel:} ratio between the 
Crab spectrum as measured by previous experiments and the PDS spectrum. 
With the exception of the Zombeck and the Bartlett measurements, all the 
other are consistent with each other within 8\%.}
\label{f:Crab}
\end{figure}

A cross--calibration of the PDS and MECS telescopes \citep[see, 
e.g.,][]{Boella97} performed with the Crab has provided a time averaged 
normalization ratio at 1 keV between PDS and MECS of $R({\rm PDS/MECS})= 
0.928\pm 0.001$ and $0.917\pm 0.001$ under the assumption of a {\sc pl} 
or a {\sc bknpl} model, respectively. When corrected upward for this 
normalization ratio, the mean value of the {\sc pl} normalization 
parameter is $N$ = $9.54\pm 0.01$, while $N = 9.39\pm 0.02$ for a {\sc bknpl}. 
A comparison of this Crab spectrum with that obtained with other 
instruments (see Fig.~\ref{f:Crab}) shows that this renormalized PDS 
spectrum is consistent within 8\% with the extrapolation of the 0.3--10 
keV spectrum obtained with the {\em XMM-Newton} EPIC--MOS camera 
\citep{Willingale01}, with the Crab spectrum obtained with {\em HEAO--1} 
A4 \citep{Jung89} and with the classical \citet{Toor74} results. In addition 
it is in excellent agreement with the mean value of Crab spectrum quoted by 
\citet{Kirsch05}. As can 
also be seen from Fig.~\ref{f:Crab}, with respect to the Crab 15--50 keV 
spectrum obtained with the PDS, that quoted by \citet{Zombeck90} is higher 
by $\sim 30$\%, while that reported by \citet{Bartlett94} is lower by 
$\sim 15$\%. In the following we assume the renormalized Crab spectrum
as our calibration scale for the CXB estimate.

\begin{figure}
\begin{center}
\includegraphics[width=\textwidth]{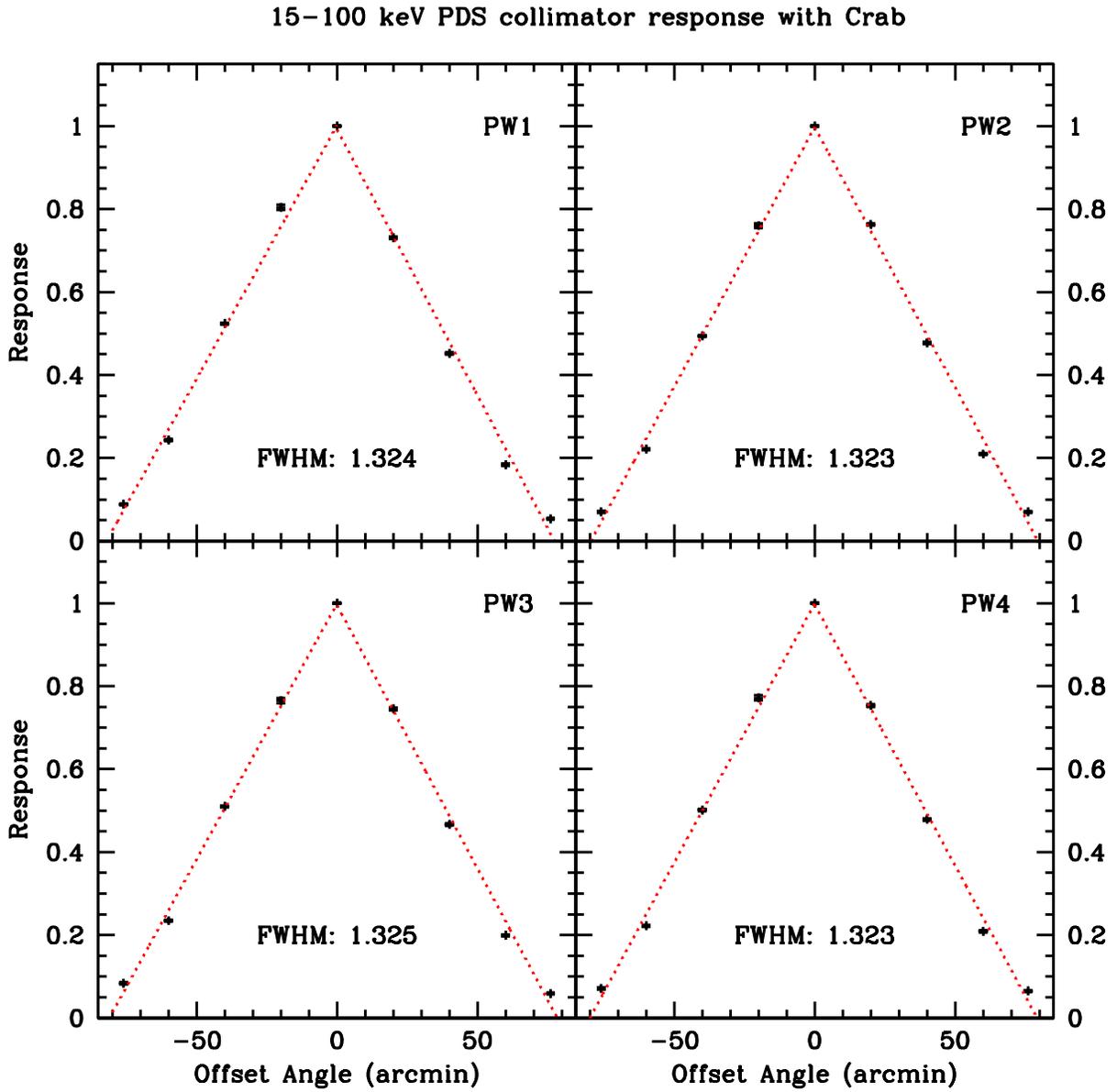}
\end{center}
\caption[]{Angular response of the PDS collimators for each phoswich unit
measured from offset observations of the Crab in 15--100 keV.}
\label{f:ang_resp}
\end{figure}

The OFF--axis response function of the PDS was tested with the Crab Nebula 
during the {\em BeppoSAX} Science Verification Phase. The Crab was 
observed in September 1996 at different offset angles with respect to the 
instrument axis and with a roll angle such as to get the narrowest angular 
response of the hexagonal collimators. Results of those measurements were 
reported \citep{Frontera97b} and now confirmed (see 
Fig.~\ref{f:ang_resp}). Up to 100 keV the angular response of each PDS 
unit to the Crab is well fit by a triangular function as expected. The derived 
FWHM is $\theta_{\rm fw} = 1\fdg32$, which is fully 
consistent with that derived in the pre-flight tests \citep{Frontera97a}. 
We also found that the response function is independent of the offset 
angle $\theta$ apart from the exposed area through the collimators, which 
linearly decreases with $\theta$. From the angular response of the PDS, we 
obtain (see Appendix A) for the geometric factor of the telescope a value $G 
= 0.295$~cm$^2$~sr, and for its solid angle a value $\Omega = 4.624\times 
10^{-4}$~sr.

\section{Measurement of the unresolved CXB}

\subsection{Adopted method}

The measurement of the unresolved CXB photon spectrum stands as one of the 
most difficult tasks in observational X--ray astronomy. Among other 
things, it requires the knowledge, for each energy channel, of the 
instrument intrinsic background count rate $\nu_{in}$ to be subtracted 
from the total background $\nu_B^{sky}$ measured during the observation of 
a blank sky field ($\nu_B^{sky} = \nu_{CXB} + \nu_{in}$, where $\nu_{CXB}$ 
is the CXB count rate entering through the telescope FOV and $\nu_{in}$ is 
the intrinsic background).  Systematic 
errors in the $\nu_{CXB}$ estimate can intervene either in a biased 
selection of blank fields or in a wrong $\nu_{in}$ evaluation or in both.

In the case of the {\em HEAO--1} A2 experiment, the total background level 
$\nu_B^{sky}$ measured during the observation of a blank sky field 
 was simultaneously observed through two collimators with 
different FOVs, one with a solid angle twice that of the other. Assuming 
$\nu_{in}$ to be independent of the instrument FOV, the difference between 
the two total background levels removes the intrinsic background and gives 
directly $\nu_{CXB}$ (M80). In the case 
of the {\em HEAO--1} A4 experiment \citep[G99,][]{Kinzer97}, the 
unresolved CXB spectrum was derived by subtracting from the $\nu_B^{sky}$, 
measured when the detectors observed blank sky fields, the background 
measured when the FOV of the detectors was shielded with a shutter made of 
a scintillator detector in anti-coincidence with the main detectors.
In both cases some  small bias in the  $\nu_{CXB}$ estimate cannot 
be excluded given that the intrinsic background is dependent on the mass 
exposed to the environmental radiation: in the first case, there are two different 
collimator apertures, while, in the second case, the shutter could modify the 
intrinsic background. 

Our measurement of the unresolved $\nu_{CXB}$ count rate is based on the 
{\em Sky-Earth Pointing} (SEP) method, in which we subtract from the 
background level $\nu_B^{sky}$ measured from a blank sky field 
($\nu_B^{sky} = \nu_{CXB} + \nu_{in}^{sky}$) the count rate level measured 
when the telescope is pointing to the dark Earth ($\nu_B^{Earth} = \nu_A + 
\nu_{in}^{Earth}$, where $\nu_A$ is the count rate due to the X--ray 
terrestrial albedo entering through the telescope FOV). The difference 
spectrum $D(E) = (\nu_{CXB} - \nu_{A}) + (\nu_{in}^{sky} - 
\nu_{in}^{Earth})$ becomes $D(E) = \nu_{CXB} - \nu_A$ if $\nu_{in}^{Earth} 
= \nu_{in}^{sky}$. Given that the radiation environment should not change 
looking to the dark Earth or to the sky, the latter condition is expected 
to be satisfied if both measurements are performed at the same cutoff 
magnetic rigidity and at the same time distance from the South Atlantic 
Geomagnetic Anomaly. The SEP strategy was also adopted for the CXB 
measurement performed with the {\em ASCA} GIS \citep{Kushino02} and {\em 
BeppoSAX} MECS \citep{Vecchi99} imaging telescopes, and with the {\em 
Rossi-XTE} PCA collimated detector \citep{Revnivtsev03}.

\subsection{Data selection}
\label{s:data}
In order to make sure that $\nu_{in}^{Earth} = \nu_{in}^{sky}$ we performed 
a careful selection of the available data. The difference $D(E)$ was 
derived only for Observation Periods (OPs) of $\ge 10$~ks duration during 
which the corresponding $\nu_B^{sky}$ and $\nu_B^{Earth}$ were measured at 
similar mean values of the cutoff rigidity. In addition, given that the 
instrument mass distribution exposed to the sky (or Earth) and thus the 
intrinsic background can change with the collimator offset angle, we 
separately derived $D(E)$ for the ON, $+$OFF and $-$OFF collimator 
positions. For the Earth pointings, we selected only those with the PDS 
axis well below the Earth limb. This method does not require a variable 
instrument configuration but requires the measurement of the albedo 
spectrum, which is not negligible at energies $>$15~keV.

In order to satisfy the blank sky field condition, in addition to 
discarding all those pointings within 15$^\circ$ from the Galactic plane, 
for the OFF-source pointings we filtered out those observations for which 
the $+$OFF and $-$OFF fields could be contaminated, e.g. from 
serendipitous X--ray sources, fast transients or solar flares. This 
selection was done by excluding from the sample those observations for 
which the difference between the two offset spectra ($+$OFF minus $-$OFF) 
were inconsistent with zero at 98\% confidence level. For the ON--source 
pointings we accepted only those fields for which the difference between 
the ON-source count rate and count rate measured at either $+$OFF and 
$-$OFF is consistent with zero within $1\sigma$, and for which a fit with 
a null constant to the difference between the corresponding offset spectra 
($+$OFF minus $-$OFF)  gives a $\chi^2$ per degree of freedom in the range 
0.8--1.2. All the sky observations were done only when the instrument axis 
pointed at a direction at least 5$^\circ$ away from the Earth limb.

As a result of the above selections, from the entire set of 868 {\em 
BeppoSAX} OPs off the Galactic plane, the number of useful OPs becomes 275 
(127 ON-source, 71 $+$OFF-source, and 77 $-$OFF--source) with a total 
exposure time of 4031 ks. The dark Earth was observed for a total of 2056 
ks.

\subsection{Results}

\begin{figure}
\begin{center}
\includegraphics[angle=270,width=\textwidth]{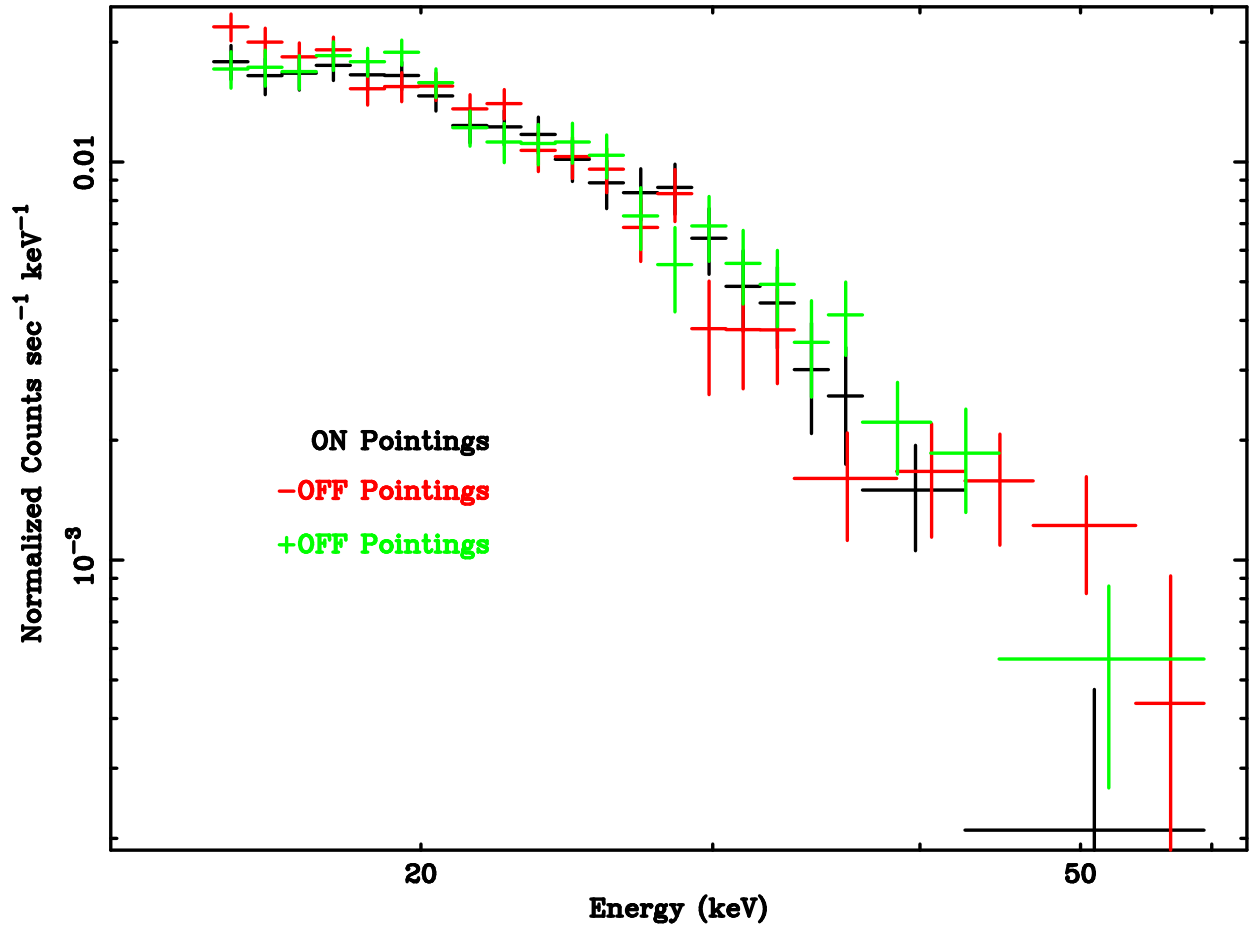}
\end{center}
\caption[]{Difference spectrum $D$ obtained from the ON-source and OFF-source
pointings.}
\label{f:D}
\end{figure}

\begin{figure}
\begin{center}
\includegraphics[angle=270,width=\textwidth]{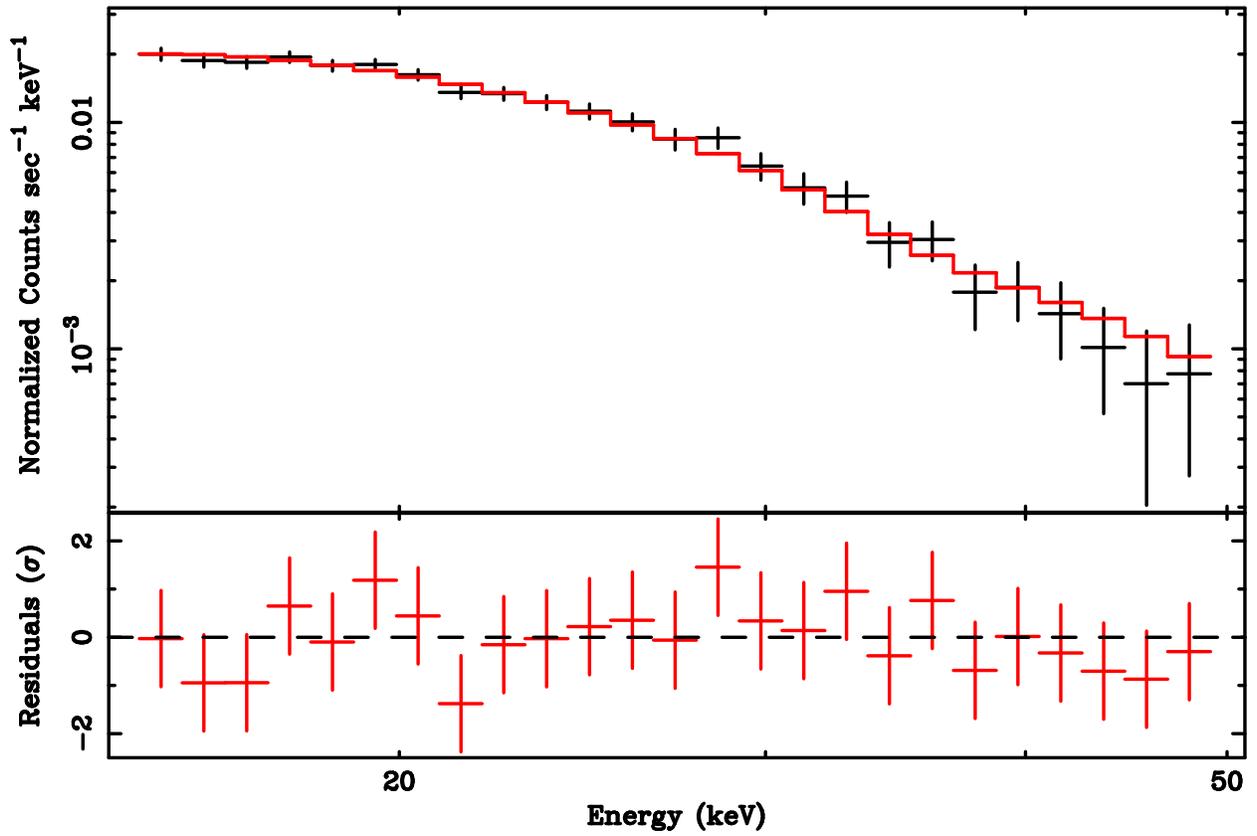}
\end{center}
\caption{{\em Top panel}: Average difference spectrum $D = \nu_{CXB} - 
\nu_A$ of all the available data, along with one of the best fit models. 
In this case the CXB spectrum is modeled with a {\sc pl} and the 
terrestrial albedo with an absorbed {\sc pl} (see text). {\em Bottom 
panel}: residuals to the model.}
\label{f:fit}
\end{figure}

We obtained three difference spectra: $D_{\rm ON}(E)$ for the 
ON--source pointings (2350 ks of exposure time), $D_{+{\rm 
OFF}}(E)$ for the $+$OFF--source pointings (800 ks) and  
$D_{-{\rm OFF}}(E)$ for the $-$OFF--source pointings (881 ks). They are 
shown in Fig.~\ref{f:D}. As can be seen, all are consistent with each 
other within their uncertainties, as also confirmed with a run test 
\citep[see, e.g.,][]{Bendat71}. The small observed deviations from each 
other give an indication of the systematic errors made in the estimate of 
the difference $\nu_{in}^{sky} - \nu_{in}^{Earth}$. However, the results 
from the offset and ON--source pointings agree within the statistical 
errors. On the basis of these results, for the derivation of the CXB 
intensity we used the sum $D(E) = D_{\rm ON}(E) + D_{+{\rm OFF}}(E) + 
D_{-{\rm OFF}}(E)$ which is well determined up to 50 keV (see 
Fig.~\ref{f:fit}).

When $D(E)$ is fit with a {\sc pl} model, we find an unacceptable 
$\chi^2/{\rm dof}$ (= 37.4/23) in the 15--50 keV band, with a
probability of 2.5\% that this high $\chi^2$ value is due to chance. 
Instead, in the 20--50 keV band, an acceptable value (= 15.5/16) is found 
with a best fit {\sc pl} photon index $\Gamma$ at 90\% confidence level 
of $3.2\pm 0.2$, much higher than that of the CXB (see 
Table~\ref{t:CXB_le}). Both results are expected if $D(E)$ does give 
the difference spectrum between the CXB and the albedo from the 
dark Earth, with the first result mainly due to the presence of a low
energy cutoff in the albedo spectrum (see Appendix B.1). The high $\Gamma$
also unequivocally shows that the albedo spectrum, above its cutoff, is harder 
than that of the CXB.
Thus, in the 15 to 50 keV band, we fit $D(E)$ with the difference of two 
model spectra, one to describe the unresolved CXB spectrum and the other 
to describe the albedo radiation spectrum.
  
For the albedo model spectrum we used a photo-electrically 
absorbed power--law $I_{\rm A}(E) = \exp{(-t_A\mu_A)} N_A 
(E/20)^{-\Gamma_A}$~photons~cm$^{-2}$~s$^ {-1}$~keV$^{-1}$~sr$^{-1}$, 
where $N_A$ is the normalization constant at 20 keV, $\mu_A$ is the air 
absorption coefficient in units of cm$^2$/g, and $t_A$ is the atmospheric 
depth that describes the well known cutoff in the albedo spectrum at $\sim 
30$~keV (see Appendix B.1). To describe the albedo low--energy cutoff, we 
developed an atmospheric absorption model within the XSPEC software 
package \citep{Arnaud96}, that we adopted for our model fitting. The model 
makes use of a grid of values of air mass X--ray attenuation coefficients 
as a function of the photon energy \citep{mass-att}. To model the CXB 
spectrum we assumed the CXB spectral shape obtained with {\em HEAO--1} 
A2$+$A4 (G99). Thus we used as input models a {\sc cutoffpl} 
and a {\sc pl} which, in the 15--50 keV interval, still gives a good description of 
this shape.

\begin{figure}
\begin{center}
\includegraphics[width=0.45\textwidth]{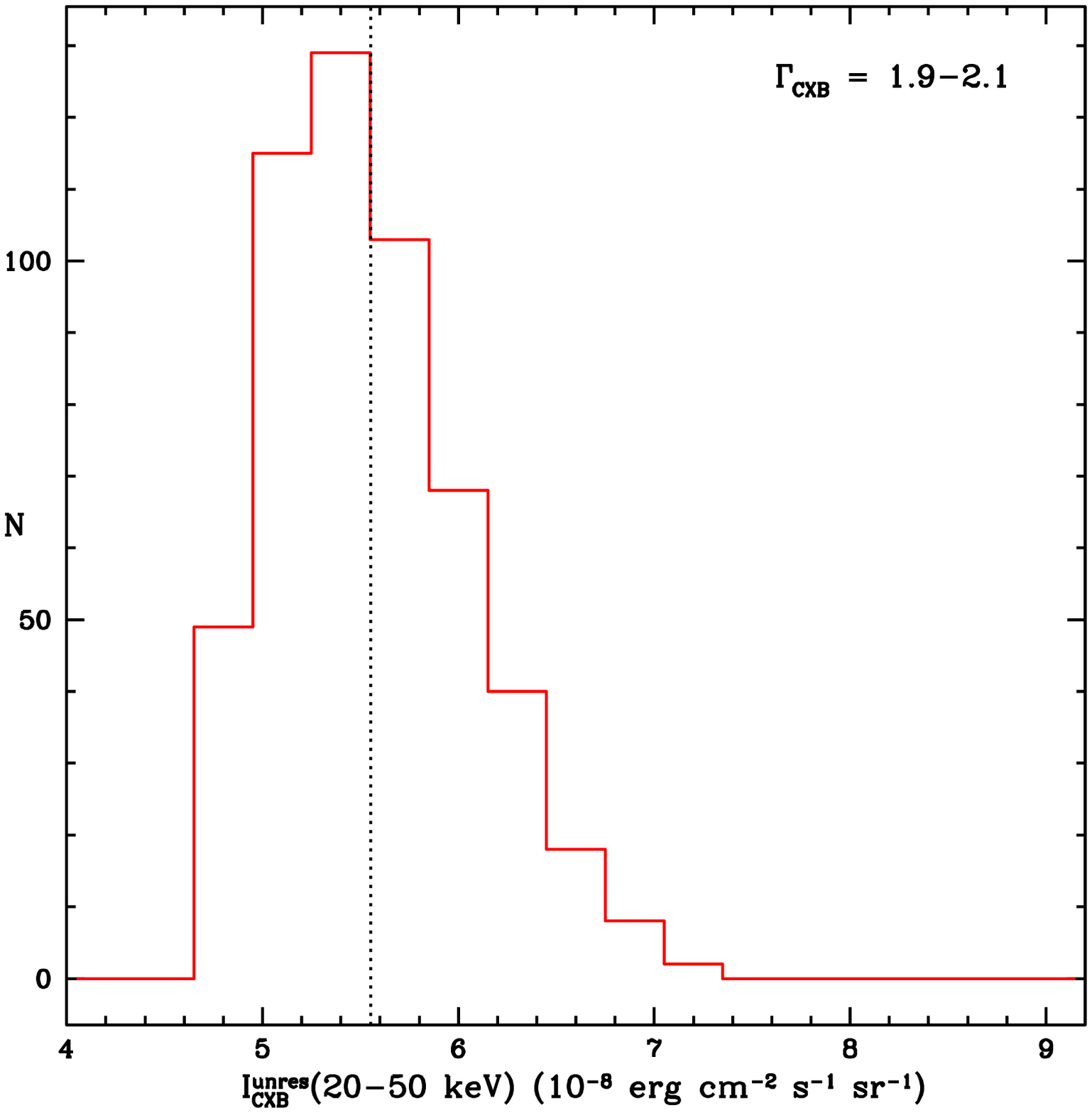}%
\includegraphics[width=0.45\textwidth]{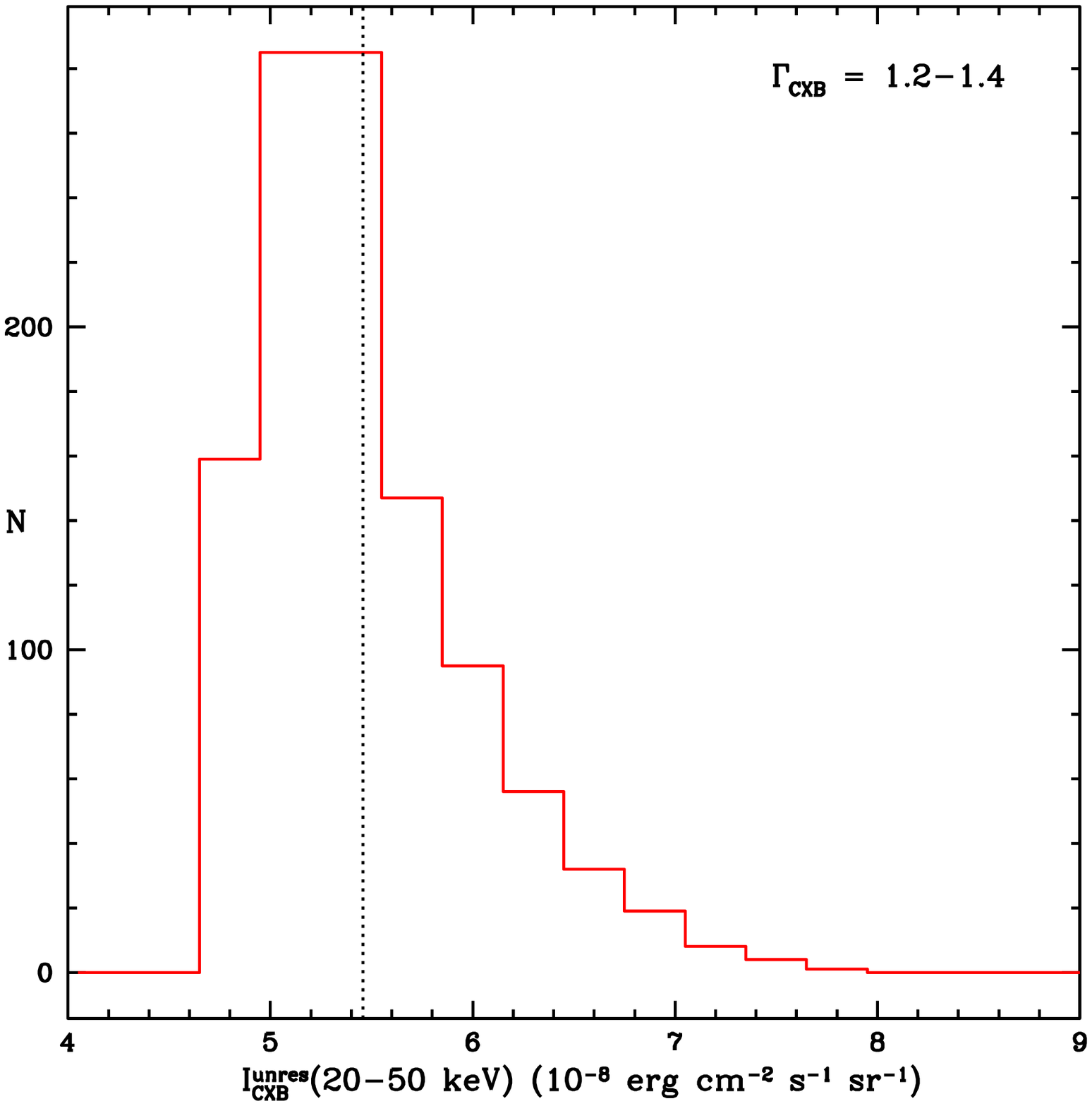}
\end{center}
\vspace*{-1cm}
\caption[]{Frequency distributions of the unresolved CXB 20--50 keV 
integrated intensity. The vertical dashed lines give the mean value of the 
distributions. {\em Left panel:}  in the case of a {\sc
pl} CXB model with $\Gamma$ in the range 1.9--2.1 (532 trials, see text). 
{\em Right panel:} in the case of a {\sc cutoffpl}
CXB model with $\Gamma$ in the range 1.2--1.4 (1091 trials, see text). 
}
\label{f:hist_fcxb}
\end{figure}

In order to better constrain the CXB spectral parameters, the fit of $D(E)$
to the data was not performed by leaving free to vary  all the model's
parameters. In return, different fits were performed, each one with  different set 
of values of the parameters frozen in the fits. These parameters were: the CXB photon 
index $\Gamma$ (when it was not left free to vary – see Tab.~\ref{t:cxb_results}), 
the atmospheric depth $t_A$ and the albedo {\sc pl} photon index $\Gamma_A$. 
The allowed ranges of $\Gamma$  were from 1.9 to 2.1 for the {\sc pl} and 
from 1.2 to 1.4 interval for the {\sc cutoffpl}, the allowed range of $t_A$ 
(in units of g~cm$^{-2}$) was from 1.4 to 6 and that of 
$\Gamma_A$ was from 1.3 to 2.0. Only the cutoff energy $E_c$ was always 
frozen to the value of 41.13 keV found with {\em HEAO--1} (see Table~\ref{t:CXB_le}), 
given that in our energy band (15--50 keV) the fits were insensitive to $E_c$. 
These ranges include the values obtained in past measurements and take into account the 
fact that the albedo mean 
slope is harder than that of the CXB, as previously discussed. In this way  
the entire parameter space was explored. By uniformly 
subdividing the allowed ranges  of $t_A$, $\Gamma_A$ and $\Gamma$ in a certain
number of subintervals, we obtained a grid of best fit values of those parameters
that were left free to vary in the fits~\footnote{In all tables, those parameters that
were frozen in the fits are shown in square brackets.}. From this grid
we have derived the frequency distribution of one of the most important quantities
that characterize our CXB estimate, i.e., the 20--50 keV integrated intensity 
\intunr\ of the unresolved CXB~\footnote{We adopt this energy band given 
that it is common to all high energy X--ray experiments quoted in 
Table~\ref{t:CXB_le} and Table~\ref{t:CXB_he}.}.  In Fig.~\ref{f:hist_fcxb} we show
this distribution, for both a {\sc pl} and {\sc cutoffpl} CXB model. 
 It was obtained by performing 1200 trials (20 steps for $t_A$, 20 for 
$\Gamma_A$ and 3 for $\Gamma$), and
accepting only those spectral solutions (532 in the case of the {\sc pl}, 1091 in the
case of the {\sc cutoffpl}) for which the {\sc pl} photon index of 
the reconstructed $D(E)$ spectrum in the 20--50 keV band is in the range 
2.9--3.5, consistently with the observed slope of $D(E)$. (Without this 
constraint the peak of the distribution occurs at slightly lower values of 
\intunr.) We verified that the moments of this distribution are independent 
of the number of chosen steps.

As can be seen, the frequency distribution of \intunr\ is well peaked 
(also the frequency distribution of the $N_{CXB}$ and $N_A$ normalizations 
and of $\Gamma$ show a 
similar shape), with a mean value slightly higher than that at the
maximum of the distribution
(see Fig.~\ref{f:hist_fcxb}). We have exploited this distribution 
to better constrain the range of the parameter values that were frozen in the 
single fits.  Having adopted the maximum likelihood method \citep{Janossy65} 
for the best estimate of the model parameters, we 
considered as best fit parameter values, reported in Table~\ref{t:cxb_results}, 
those that, in the narrow flat top region of the frequency distribution around the 
mean value, give the minimum $\chi^2$. 
In Fig.~\ref{f:fit} we show the best fitting curve to 
$D(E)$ and the corresponding residuals in the case of a {\sc pl} as input 
model for the unresolved CXB, and dark Earth albedo spectral parameters frozen at
the values shown  in Table~\ref{t:cxb_results}.  

Using the \intunr\ frequency distributions, we also derived the upper limit 
to the unresolved CXB estimate finding that, independently of the 
CXB input model, 90\% of the data points have $\intunr < 6.5\times 
10^{-8}$~erg~cm$^{-2}$~s$^{-1}$~sr$^{-1}$. We assume this value as upper 
limit to our estimate of the unresolved CXB. We notice that, by increasing 
the range of the CXB photon index values to 1.8--2.2 for the {\sc pl}, and 
to 1.1--1.5 for {\sc cutoffpl}, we find similar \intunr\
distributions with a mean at slightly lower intensity values but with 
upper limit almost unchanged with respect to that given above.

In Appendix B.2 we report and discuss the obtained results on the terrestrial 
albedo from the dark Earth. For a comparison of the CXB spectrum with the
albedo spectrum see Fig.~\ref{f:albedo}.

\section{The total (resolved plus unresolved) CXB \label{s:total}}

Exploiting the PDS pointings, we also performed an estimate of the 
contribution of resolved sources to the 15--50 keV integrated CXB 
intensity.  This estimate could not be done using the ON-source pointings 
of the PDS, given that most of them were pointed observations of specific 
targets by the {\em BeppoSAX} Narrow Field Instruments LECS, MECS, 
HPGSPC, and PDS. However, 33 of the PDS OFF--source fields, that were excluded
from the data set for the unresolved CXB intensity determination 
(see Sect.~\ref{s:data}), showed significant count excesses (2.5--7$\sigma$), 
consistent with the presence of serendipitous X--ray sources. 
The spectra of these excesses, 
when fit with a {\sc pl} model, gave a weighted mean value of their photon 
indices equal to $1.65\pm0.20$, while their intensity gave a 15--100 keV 
energy flux in the range (1.4--20)$\times 10^{-12}$~erg~cm$^{-2}$~s$^{-1}$. 
A search of their counterparts is now in progress. Their contribution 
increases the CXB intensity by 4.7\%. Adding the contribution of brighter 
sources \citep[see, e.g.,][]{Krivonos05} does not significantly change 
this figure.

\begin{figure}
\begin{center}
\vspace*{-2.5cm}
\includegraphics[angle=270,width=0.9\textwidth]{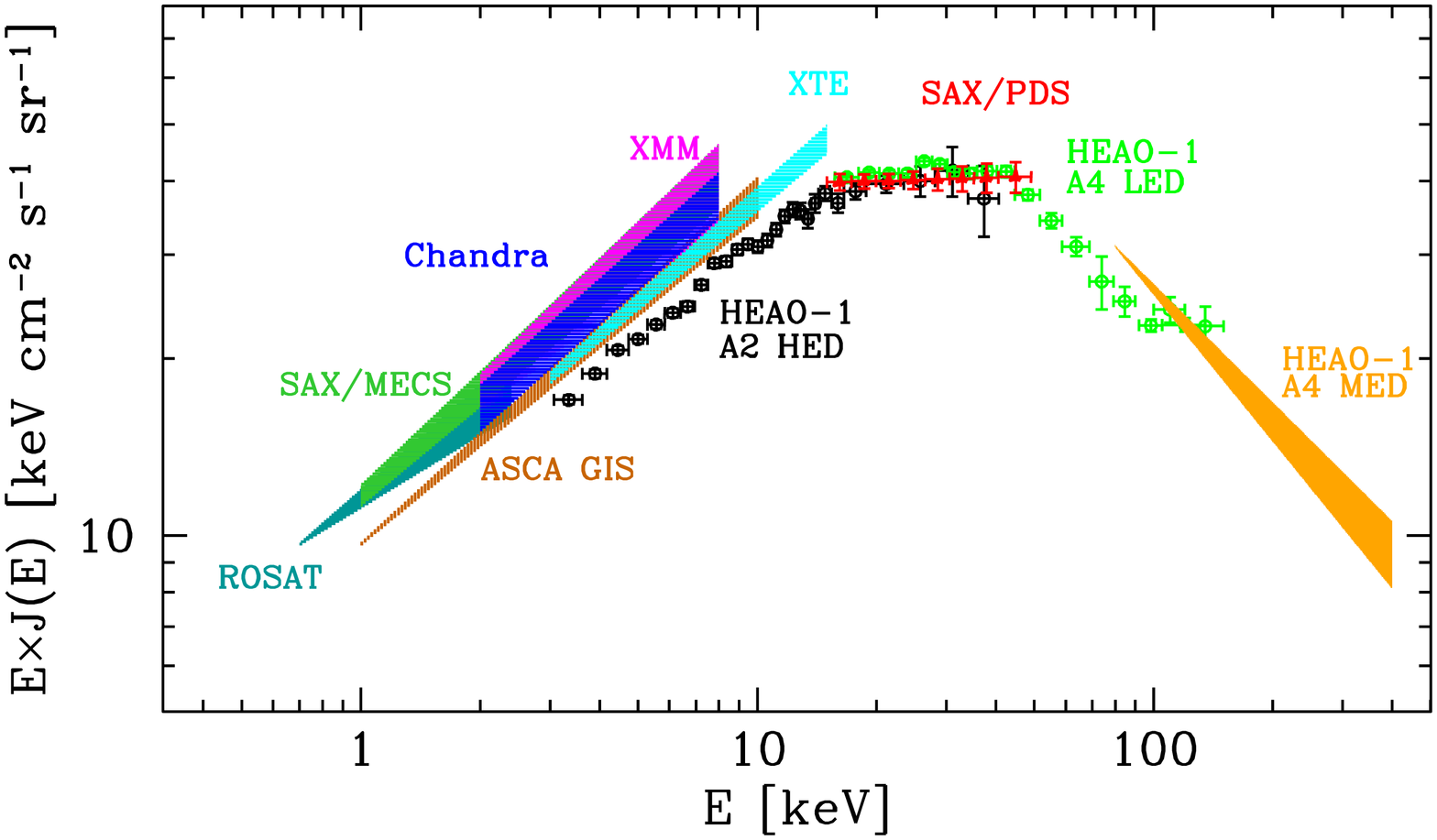}
\strut\vspace*{-1.5cm}
\includegraphics[angle=270,width=0.9\textwidth]{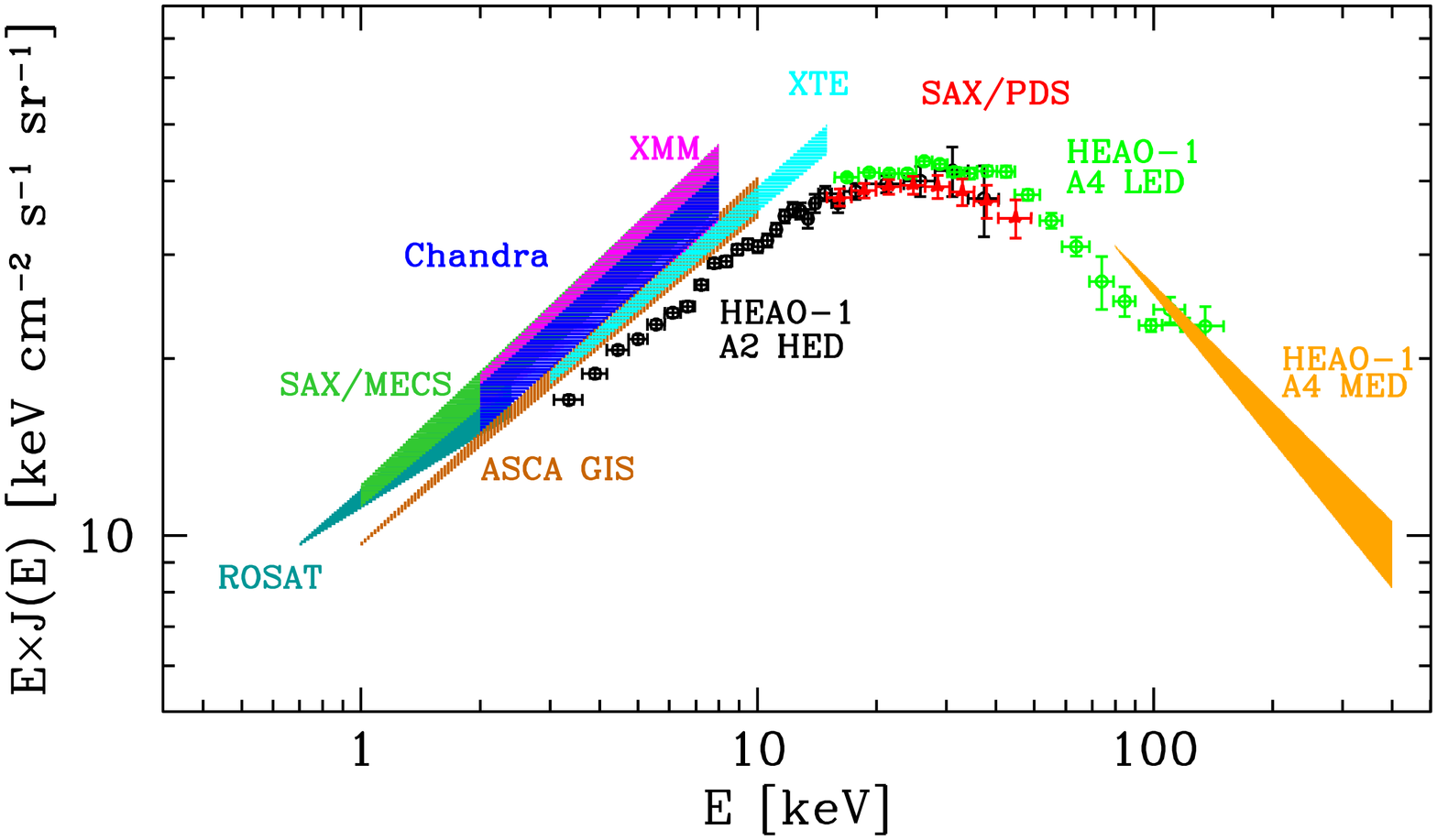}
\end{center}
\vspace*{-1cm}
\caption[]{Total (unresolved plus resolved) $E\,J(E)$ CXB spectrum as 
observed with the PDS experiment (red points) compared with measurement 
results obtained with other missions. {\em Upper panel}: The energy 
spectrum $J(E)$ modeled with a {\sc pl} with $\Gamma= 1.98$ (see 
Table~\ref{t:cxb_results}). {\em Bottom panel}: The energy spectrum $J(E)$ 
modeled with a {\sc cutoffpl} with $\Gamma= 1.4$ (see 
Table~\ref{t:cxb_results}) and $E_{\rm c}$ fixed at the value of 41.13 keV 
obtained with {\em HEAO--1\/} A2$+$A4 (G99).}
\label{f:nuFvu}
\end{figure}

Taking into account these results, in Table~\ref{t:CXB_he} we report, 
for each of the used input models, 
the normalization $N_{\rm CXB}^{\rm tot}$ of the total CXB (unresolved
plus resolved) spectrum, while in 
Fig.~\ref{f:nuFvu} we show, for the {\sc pl} and {\sc cutoffpl} models, 
the best fit $E\,J(E)$ spectrum of the total CXB. In  Table~\ref{t:CXB_he} 
and Fig.~\ref{f:nuFvu} we also compare our measurement with the past results.

\section{Discussion}

It is apparent from Fig.~\ref{f:nuFvu} and Table~\ref{t:CXB_he} 
that our best fit \inttot\ is in excellent agreement with 
that obtained with {\em HEAO--1\/} A2 (M80), and slightly lower (from 3 to 
10\%, depending on the input model) than that quoted by G99. (As
discussed in Section~\ref{s:cal}, the use of our flux scale 
calibration is more realistic; had we used the scale calibration given by \cite{Jung89} for
{\em HEAO--1} A4, our best fit \inttot\ would range from 0.95 to 1.03 times 
the corresponding value derived from G99.) The best 
fit value of the maximum CXB flux density is obtained
in the 26--28 keV band and ranges from 39.4 to 40.2~keV~(cm$^2$~s~sr)$^{-1}$ 
depending on the model assumed, with a statistical uncertainty in the 
centroid of $\pm1.5$ keV~(cm$^2$~s~sr)$^{-1}$ at 90\% confidence level
for a single interesting parameter.

We have also evaluated the upper limit to the CXB intensity that can be 
marginally accommodated by our data, by exploring the space of all the 
parameters involved in the fits. We find that, taking also into account
the contribution of the resolved sources, independently of the CXB 
model, in 90\% of this multi-parameter space the \inttot\
is lower than $6.8\times 
10^{-8}$~erg~cm$^{-2}$~s$^{-1}$~sr$^{-1}$, which is 12\% higher than the
best fit CXB intensity value quoted by G99 and 21\% higher than that 
quoted by M80 (see Table~\ref{t:CXB_he}).

\begin{figure}
\begin{center}
\includegraphics[angle=270,width=\textwidth]{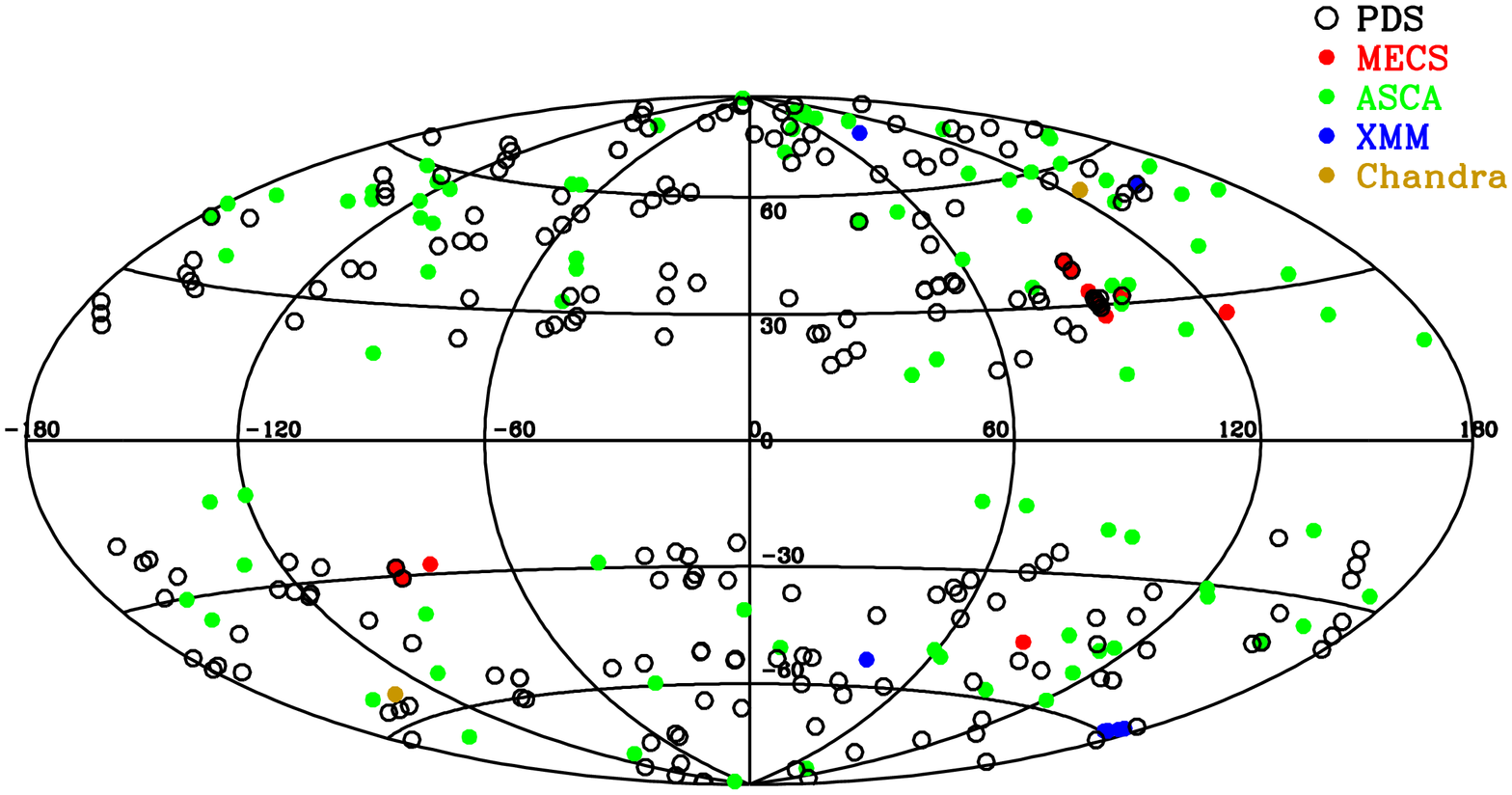}
\end{center}
\caption[]{Sky covered by the PDS compared with that covered by different 
focusing missions for the unresolved CXB determination. The 172 PDS 
pointings correspond to 265 deg$^2$. Sky coverage of focusing telescopes: 
0.73 deg$^2$ \citep[{\em MECS\/};][]{Vecchi99}; 50 deg$^2$ \citep[{\em 
ASCA\/};][]{Kushino02}; 1.2 deg$^2$ \citep[{\em XMM-Newton\/};][]{Lumb02}; 
0.5 deg$^2$ \citep[{\em Chandra\/};][]{Hickox06}. For the 34 {\em 
XMM-Newton} pointings used by \citet{Deluca04}, coverage is 5.5 deg$^2$, 
but coordinates are not available.}
\label{f:sky_coverage}
\end{figure}

Even this upper limit disagrees with the extrapolation to higher energies 
of the low energy ($<$10 keV) CXB estimates obtained with the focusing 
telescopes aboard {\em BeppoSAX} \citep{Vecchi99}, {\em XMM--Newton} 
\citep{Lumb02,Deluca04}, and {\em Chandra} \citep{Hickox06}, but it is 
consistent with those obtained with {\em ASCA} and {\em RXTE}.
Thus, if we exclude a change 
in the CXB spectral shape derived with {\em HEAO--1}, our results raise 
the issue about the origin of the highest CXB intensities being quoted at 
lower energies. Differences in the flux scale calibration do not appear to 
be the origin of these discrepancies, as discussed in  Section~\ref{s:cal}. 
One may think that part of the 
discrepant results could be due to the amount of sky solid angle surveyed, 
which is very large in the case of {\em HEAO--1} and {\em BeppoSAX} PDS, 
and very small in the case of {\em BeppoSAX} MECS, {\em XMM--Newton} and 
{\em Chandra} (see Fig.~\ref{f:sky_coverage}), although, as also discussed 
by \citet{Barcons00}, this would imply that the sky regions surveyed by 
these telescopes are systematically brighter than the average sky sampled 
with {\em HEAO--1} and {\em BeppoSAX} PDS. Another possible origin of the 
highest 2--10 keV CXB estimates could be due to systematic errors in the 
response function used for the diffuse emission (e.g., an underestimate of 
the stray light). For instance, in the case of MECS this function 
could be well tested 
and cross-calibrated with the PDS only for point-like sources.

Independently of the CXB intensity issue at lower energies, our 
observational findings bear at least two important astrophysical 
consequences. Firstly, they provide a robust estimate of the accretion 
driven power integrated over cosmic time, including that produced by the 
most obscured AGNs. The AGN synthesis models, tuned to attain the PDS CXB 
level and to account for the hard X--ray spectral shape 
\citep{Lafranca05,Gilli06}, predict that at the bright fluxes ($>10^{-12}$ 
erg~cm$^{-2}$~s$^{-1}$) reachable by the PDS observations and by coded 
mask instruments like {\em INTEGRAL} IBIS and {\em Swift} BAT, the 
detectable fraction of extremely obscured AGNs (Compton thick, $N_{\rm 
H}>10^{24}$ cm$^{-2}$) is $\sim$10\%, and it increases to 20--25\% at the 
fluxes reachable by focusing telescopes \citep[e.g.][]{Ferrando06}. It 
should be noted that the highest CXB intensities claimed at low energies 
\citep[ see Table~\ref{t:CXB_le}]{Vecchi99,Deluca04,Hickox06} would entail 
a much larger number (a factor 2--3) of Compton thick AGNs 
\citep{Gilli06}, a prediction barely consistent with the present 
observational evidence. Our result implies a present black hole mass 
density of $\sim 3\times10^5$~M$_\odot$~Mpc$^{-3}$, using an admittedly 
uncertain bolometric correction of 30 for the 15--50 keV band and an 
efficiency of 0.1 in converting gravitational into radiation energy. This 
corresponds to a fraction of $6\times 10^{-5}$ of all baryons being locked 
into the supermassive black holes, using a cosmic baryon density of $\sim 
4 \times 10^{-31}$ g~cm$^{-3}$ in agreement with the ``concordance'' 
cosmology model.

Secondly, under the assumption that the {\em HEAO--1} spectral shape 
(G99) applies down to 2 keV, we find that the summed 
contribution of the observed X--ray source counts in the 2--10 keV band 
\citep{Moretti03} exceeds the PDS CXB best fit level by $\sim$11\%. This 
apparent contradiction vanishes if one takes into account the above 
discussed upper limit in the PDS CXB intensity level and the error 
($\pm$7\%) associated with the source count evaluation \citep{Moretti03}. 
As a consequence, our measurement suggests that it is quite possible that 
almost all the CXB in the 3--8 keV band has already been resolved into 
sources down to the faintest fluxes of the {\em Chandra} deep fields. Any 
substantial contribution to the CXB from other classes of sources and 
diffuse WHIM should be confined at photon energies below $\sim$3 keV, as 
it has already been indicated in the case of star forming galaxies 
\citep{Ranalli03}.

\acknowledgments

We wish to thank L.~Bassani, A.~Fabian and G.C. Perola for useful suggestions.
We wish to acknowledge the several comments of the anonymous referee that helped
us to make the paper clearer and stronger.  
{\em BeppoSAX} was a joint program of the Italian Space Agency (ASI) and the 
Netherlands Agency for Aerospace Programs. This research was supported by 
ASI and Ministry of Education, University and Research of Italy (COFIN 
2002 and 2004).

\clearpage

\appendix

\section{The solid angle of the PDS instrument}

The solid angle of the PDS is derived from the Geometric Factor of the 
telescope defined as \citep{Peterson73,Horstman75}
\begin{equation}
G = \int_{\Omega} A(\theta) d\Omega,
\end{equation}
\noindent where $A(\theta)$ is the exposed detector geometric area through 
the collimators at an offset angle $\theta$, and $\Omega$ is the solid 
angle of the instrument Field of View (FOV). The expression of $G$, for an 
hexagonal collimator like that of the PDS, can be found in 
\citet{Horstman75} and is given by
\begin{equation}
G = \pi A(0) \tan^2 (0.5 \tan^{-1} \frac{0.93939d_m}{h})
\end{equation}
\noindent where $A(0)$ is the ON--axis geometric area through the 
collimator (640 cm$^2$), $d_m$ is the diameter of the circumscribed circle 
to the hexagonal collimator cells, and $h$ is collimator height. The ratio 
$d_m/h$ is related to the minimum FWHM of the angular response of the PDS 
$\theta_{\rm fw}$ through the relation
\begin{equation}
\frac{d_m}{h} =  \frac{2}{\sqrt{3}}\tan{\theta_{\rm fw}}
\end{equation}
Using the value of $\theta_{\rm fw}$ derived from the offset Crab 
observations, we obtain a value of $G = 0.295$~cm$^2$~sr and, dividing by 
$A(0) = 640$~cm$^2$, we find the telescope solid angle $\Omega = 
4.624\times 10^{-4}$~sr.

\section{The albedo spectrum\label{s:albedo}}

\subsection{The past measurements}

The terrestrial gamma--ray albedo radiation is mainly the result of the 
interactions with the upper atmosphere of the Cosmic Rays and, at X--ray 
energies, of the CXB and of the discrete X--ray source radiation (Compton 
reflection). It was investigated since the late 1960s \citep[see][ for a 
review]{Peterson75}. Most of the properties of the atmospheric gamma--rays 
were obtained with balloon experiments, mainly launched from Palestine 
(Texas, USA). Empirical models of the atmospheric gamma--ray emission, 
based on observational results, were worked out by various authors 
\citep{Peterson75,Ling75,Dean89}. From these observations, it can be seen 
that the gamma--ray emission properties depend on various parameters, like 
the geomagnetic latitude, the energy band, the direction of emission 
(downward, upward) and the altitude from the Earth. The atmospheric 
radiation measured by a satellite depends on these parameters, but, unlike 
the radiation observed by a balloon experiment, a satellite observes only 
the radiation emerging from the top of the atmosphere (albedo radiation). 
Thus its spectral properties, specially at low energies ($>$10 keV), are 
expected to be different from those observed inside the atmosphere. 
Measurements of the X--ray albedo radiation are reported in 
\citet{Schwartz74a} for photon energies in the 10--300 keV band, and in 
\citet{Imhof76} above 40 keV, while for energies higher than 150 keV see, 
e.g., \citet{Letaw86} and \citet{Dean89}.

The albedo spectrum of \citet{Schwartz74a}, which was obtained with a 
0.5~cm scintillator detector and a wide FOV ($23^\circ$ at zero response) 
aboard the {\em OSO--3} satellite in a nearly circular orbit at an 
inclination of 33$^\circ$ and an altitude of 550 km \citep{Schwartz70}, 
above 40 keV is consistent with a {\sc pl} (see Fig.~\ref{f:albedo}), 
while below 40 keV it shows a flattening with a definitive low energy 
cutoff below 30 keV. This cutoff is also observed with balloon experiments 
\citep{Peterson73,Peterson75,Schwartz74a} and it is attributed to 
self-absorption of the radiation collectively emitted from different atmospheric 
layers. The {\sc pl} model above 40 keV is confirmed by the albedo 
spectrum measured with a 50 cm$^3$ Ge(Li) cooled detector with a wide FOV 
as well ($\pm45^\circ$ at zero response) aboard the low--altitude 
polar--orbiting satellite {\em 1972--076B} \citep[see 
Fig.~\ref{f:albedo};][]{Imhof76}.

Following \citet{Schwartz69}, the best fit to the $>$10 keV albedo 
spectrum measured with {\em OSO--3} is obtained with a photo-electrically 
absorbed {\sc pl} $I_{\rm A}(E) = \exp{(-t_A\mu_A)} N_A 
(E/20)^{-\Gamma_A}$~photons~cm$^{-2}$~s$^ {-1}$~keV$^{-1}$~sr$^{-1}$, with 
photon index $\Gamma_A = 1.7\pm 0.3$, and atmospheric thickness $t_A = 
1.75\pm 0.15$ g/cm$^2$ (at 90\% confidence level). In the case of the 
polar--orbiting satellite, \citet{Imhof76} found that above 40 keV the 
photon spectrum is consistent with a {\sc pl} with index ranging from 
$\sim$1.34 to $\sim$1.39, depending on the latitude scanned.

On the basis of these observations, we have assumed a 
photo-electrically absorbed {\sc pl} as a model spectrum for the albedo 
radiation from the dark Earth.

\subsection{Our results}

\begin{figure}
\begin{center}
\includegraphics[width=\textwidth]{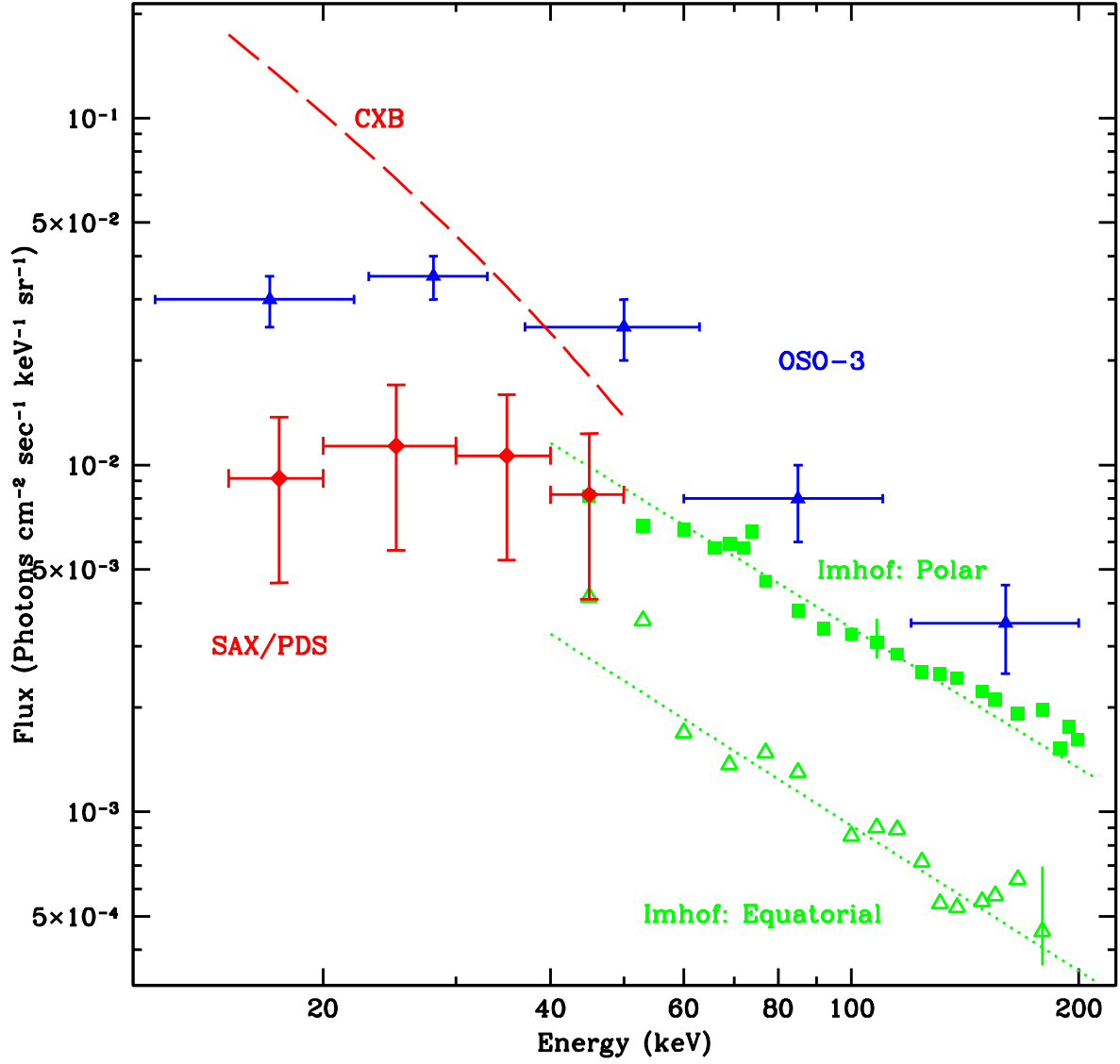}
\end{center}
\vspace*{-1.5cm}
\caption[]{Comparison of the best fit photon spectrum of the dark terrestrial
albedo as derived by  the PDS measurement (red points) with the results found 
by \citet[ {\em OSO--3} satellite, blue points]{Schwartz74a} and by \citet[ {\em 
1972--076B} satellite, green points]{Imhof76}. Also the best fit 
CXB photon spectrum assuming as input model a {\sc cutoffpl} with $\Gamma = 1.4$ (see Table~\ref{t:cxb_results})
is shown for comparison with the derived albedo spectrum.}
\label{f:albedo}
\end{figure}

In addition to the parameter values reported in Table~\ref{t:cxb_results}, 
we show in Fig.~\ref{f:albedo} the derived spectrum of the terrestrial 
albedo from the dark Earth, compared with that of the CXB. 
As can be seen, the derived albedo spectrum is located between the {\em 
OSO-3} results and the albedo spectrum derived by \citet{Imhof76}, with a 
20--50 keV integrated intensity of $(8\pm 2)\times 
10^{-10}$~erg~cm$^{-2}$~s$^{-1}$~sr$^{-1}$. It should be noticed that the 
intensity of the albedo radiation depends on the CR flux hitting the 
Earth, and thus on magnetic latitude. Given that in different Earth 
pointings we pointed to the Earth along generally different directions and 
thus to different magnetic latitudes, the derived spectrum is latitude 
averaged, as partially done also in the case of the {\em OSO--3} and {\em 
1972--076B} satellites due to their wide FOVs. We also notice that at 
different latitudes we do not observe the upward albedo, but the albedo 
emerging at different zenith angles $Z$. However, as discussed by 
\citet{Ling75}, the atmospheric spectrum at low energies is expected to be 
not strongly anisotropic with $Z$.

\clearpage

\bibliographystyle{apj}
\bibliography{apj-jour,cxb_biblio}


\begin{deluxetable}{ll|cccccc}
\rotate
\tabletypesize{\scriptsize}
\tablecaption{Summary of the CXB past measurements compared with {\em HEAO--1} (G99).
\label{t:CXB_le}}
\tablecomments{The reported energy band gives the interval in which the CXB
spectrum and its model parameters  have been determined.  
$R_{\rm HEAO-1}^{2-10\;{\rm keV}}$ gives the ratio between 
the  2--10 keV intensity $I(2-10\;{\rm keV})$ estimated from the reported
parameters and that obtained with {\em HEAO--1} 
($I_{HEAO-1}(2-10 keV) = (5.41\pm 0.08)\times 
10^{-8}$~erg~cm$^{-2}$~s$^{-1}$~sr$^{-1}$). Likewise $R_{\rm HEAO-1}^
{20-50\;{\rm keV}}$ gives the ratio between the estimated 20--50 keV intensity 
$I(20-50\;{\rm keV})$ and that obtained with {\em HEAO--1}($I_{HEAO-1}
(20-50\;{\rm keV}) = (6.06\pm 0.06)\times 10^{-8}$~erg~cm$^{-2}$~s$^{-1}$~sr$^{-1}$).
In square parenthesis the parameter values that were kept constant 
by the quoted authors. 
Uncertainties are 1$\sigma$ errors for a single parameter. Where not reported, 
the uncertainties are very small or are not reported in the quoted papers.}
\tablewidth{0pt}
\tablehead{
\colhead{Instrument} & \colhead{Ref.} & \colhead{Energy band} &
\colhead{Model} & \colhead{$\Gamma$} & \colhead{$kT$ or $E_c$} & 
\colhead{$R_{\rm HEAO-1}^{2-10\;{\rm keV}}$} & 
\colhead{$R_{\rm HEAO-1}^{20-50\;{\rm keV}}$} \\
\colhead{} & \colhead{} & \colhead{(keV)} & \colhead{} & \colhead{} & 
\colhead{(keV)} & \colhead{} & \colhead{} \\
}
\startdata
Composite       & (1)  & 1--20   & PL$^a$       & $1.59 \pm 0.02$   & --            
& 1.07           & --  \\
Composite       & (1)  & 20--200 & PL$^a$       & $2.040 \pm 0.013$ & --            
& --             & $0.86\pm 0.05$ \\
Composite       & (2)  & 20--165 & PL$^a$       & $2.17 \pm 0.07$   & --            
& --             & $0.90\pm 0.18$  \\
HEAO--1/A2      & (3)  & 3--50   & BREMSS$^b$   & --                & [40]          
& $0.98\pm 0.10$ & $0.92\pm 0.05$   \\
HEAO--1/A2      & (4)  & 2--10   & PL$^a$       & [1.4]             & --            
& $1.19\pm 0.06$ & -- \\
HEAO--1/A2      & (5)  & 2--10   & PL$^a$       & [1.558]$^c$       & --            
& $1.05\pm 0.06$ & --  \\
HEAO--1/A2$+$A4 & (6,7)& 3--60   & CUTOFFPL$^d$ & $1.29\pm 0.02$    & $41.13\pm0.62$
&  1             & 1 \\
Rocket          & (8)  & 2--6    & PL$^a$       & [1.4]             & --            
& $1.34\pm 0.21$ & -- \\
ROSAT/PSPC      & (9)  & 0.7--2.4& PL$^a$       & $1.50 \pm 0.09$   & --            
& $1.20\pm 0.05$ & -- \\
SAX/MECS        & (10) & 1--8    & PL$^a$       & $1.40 \pm 0.04$   & --            
& $1.43\pm 0.08$ & -- \\
ASCA/SIS        & (11) & 1--7    & PL$^a$       & $1.41 \pm 0.03$   & --            
& $1.06\pm 0.05$ & -- \\
ASCA/GIS        & (12) & 1--10   & PL$^a$       & [1.4]             & --            
& $1.18\pm 0.02$ & -- \\
XMM/EPIC-MOS/PN & (13) & 2--8    & PL$^a$       & $1.42 \pm 0.03$   & --            
& $1.30\pm 0.14$ & -- \\
RXTE/PCA        & (14) & 3--20   & PL$^a$       & $1.42 \pm 0.02$   & --            
& $1.15\pm 0.02$ & -- \\
XMM/EPIC-MOS    & (15) & 2-8     & PL$^a$       & $1.41 \pm 0.06$   & --            
& $1.36\pm 0.10$ & -- \\
Chandra/ACIS-I  & (16) & 2--8    & PL$^a$       & [1.4]             & --            
& $1.33\pm 0.13$ & -- \\
\enddata

\tablerefs{(1) \citet{Horstman75}; (2) \citet{Kinzer78}; (3) 
M80; (4) \citet{Revnivtsev05}; (5)\citet{Jahoda06}; (6) 
\citet{Gruber92}; (7) G99; (8) \citet{McCammon83}; (9) 
\citet{Georgantopoulos96}; (10) \citet{Vecchi99}; (11)  
\citet{Gendreau95}; (12) \citet{Kushino02}; (13) \citet{Lumb02}; (14) 
\citet{Revnivtsev03}; (15)  \citet{Deluca04}; (16) \citet{Hickox06}.}
\tablenotetext{a}{PL model: $I(E)\propto E^{-\Gamma}$
 ph~cm$^{-2}$~s$^{-1}$~keV$^{-1}$~sr$^{-1}$}
\tablenotetext{b}{Bremsstrahlung as described in the XSPEC user manual 
where $kT$ is the plasma temperature.}
\tablenotetext{c}{The best fit was obtained with two {\sc pl} plus a {\sc 
cutoffpl} plus two edges.}
\tablenotetext{d}{$I(E) \propto E^{-\Gamma}
 \exp{(-E/E_c)}$~photons~~cm$^{-2}$~s$^{-1}$~keV$^{-1}$~sr$^{-1}$}

\end{deluxetable}

\newpage

\begin{deluxetable}{l|cccccccc}
\tabletypesize{\scriptsize}
\rotate
\tablecaption{Spectral parameters of the unresolved
CXB and of the dark Earth albedo as derived from the PDS measurement. 
\label{t:cxb_results}}
\tablecomments{$N_{\rm CXB}^{\rm unres}$ is the unresolved CXB normalization at 
20 keV, while $I_{\rm A}({\rm 20\;keV})$ gives the intensity of the dark Earth 
albedo at 20 keV.  The last column gives the 
unresolved 20--50 keV integrated intensity. The parameters that were frozen in 
the fits are shown in square brackets. The quoted uncertainties are errors at 90\% 
confidence level for a single parameter.}
\tablewidth{0pt}
\tablehead{
\colhead{CXB model} & \colhead{$N_{\rm CXB}^{\rm unres}\ ^c$}  &  
\colhead{$\Gamma$} & \colhead{$E_{\rm c}\ ^d$} & 
\colhead{$I_{\rm A}({\rm 20\;keV})\ ^c$}  & \colhead{$\Gamma_A$} & 
\colhead{$t_A\ ^e$} & \colhead{$\chi^2/{\rm dof}$} & 
\colhead{\intunr\ $^f$} \\
}
\startdata
Power-law$^a$        & $0.096\pm 0.003$  & [1.98] & --      &
$0.014\pm 0.002$ & [1.38] & [1.91] & 9.43/23 & $5.62\pm 0.29$\\ 
\hline
Cutoff power-law$^b$ & $0.151\pm 0.005$ & [1.4]  & [41.13] &
$0.011\pm 0.001$ & [1.53] & [2.42] &  9.2/23 & $5.27\pm 0.29$ \\ 
\hline
Cutoff power-law$^b$ & $0.160^{+0.037}_{-0.016}$  &
$1.4_{-0.6}^{+0.4}$ & [41.13] &  $0.011_{-0.006}^{+0.015}$ & [1.3] &  [1.4] & 
9.0/22 & $5.62 \pm 0.30$ \\ 
\hline
Cutoff power-law$^b$ & $0.141\pm 0.004$  & [1.29] & [41.13] & 
$0.0040\pm 0.0004$ & [2.0] & [4.98] & 9.2/23 & $5.18\pm 0.27$ \\ 
\hline
\enddata
\tablenotetext{a}{$I_{\rm CXB}(E) = N_{\rm CXB} (E/20)^{-\Gamma}$}
\tablenotetext{b}{$I_{\rm CXB}(E) = N_{\rm CXB} (E/20)^{-\Gamma} \exp{(-E/E_c)}$}
\tablenotetext{c}{In units of photons~cm$^{-2}$~s$^{-1}$~keV$^{-1}$~sr$^{-1}$}
\tablenotetext{d}{In units of keV}
\tablenotetext{e}{In units of g~cm$^{-2}$}
\tablenotetext{f}{In units of $10^{-8}$~erg~cm$^{-2}$~s$^{-1}$~sr$^{-1}$.}

\end{deluxetable}

\newpage

\begin{deluxetable}{ll|ccccccc}
\tabletypesize{\scriptsize}
\rotate
\tablecaption{PDS  results on the total (resolved plus unresolved) 
CXB compared to the past measurements.\label{t:CXB_he}}
\tablecomments{For each model, the energy band of the CXB spectral determination,
the parameters of its photon spectrum and the  20--50 keV energy flux per 
steradian are reported. 
For flux scale calibration purposes, when available, also the 20--50 Crab 
flux predicted from the single experiments is reported. Uncertainties are 
1$\sigma$ errors. Parameters in square parenthesis are those kept fixed in 
the single fits.}
\tablewidth{0pt}
\tablehead{
\colhead{Experiment} & \colhead{Ref.} & \colhead{Energy band} &
\colhead{Model} & \colhead{$N$}  & \colhead{$\Gamma$} &  
\colhead{$kT$ or $E_c$} & \colhead{\inttot\ $^f$} & 
\colhead{$F_{\rm Crab}(20-50\;{\rm keV})\ ^g$} \\
\colhead{} & \colhead{} & \colhead{(keV)} & \colhead{} & \colhead{} & \colhead{} &
\colhead{(keV)} \\
}
\startdata
Composite     & (1) & 20--200 & PL$^a$       & $40.7\pm 2.3$    & $2.040\pm 0.013$ &
--      & $5.2\pm 0.3$   & -- \\
Balloon       & (2) & 20--165 & PL$^a$       & $67\pm 13$       & $2.17\pm 0.07$   &
--      & $5.47\pm1.06$  & -- \\
HEAO--1/A2    & (3) & 3--50   & BREMSS$^b$   & $13.95\pm 0.70$  &  --              &
[40]    & $5.6\pm 0.3$   & -- \\
HEAO--1/A2+A4 & (4) & 3--60   & CUTOFFPL$^c$ & $7.877\pm 0.08$  & [1.29]           &
[41.13] & $6.06\pm 0.06$ & $9.83\pm 0.03$ \\
SAX/PDS       & this paper & 15--50 & PL$^d$ & $0.100\pm 0.002$ & [1.98]           &
--      & $5.89\pm 0.19$ & $9.22\pm 0.01$ \\
SAX/PDS       & this paper & 15--50 & CUTOFFPL$^e$ & $0.158\pm 0.03$ & [1.4]       &
[41.13] & $5.52\pm 0.18$ & $9.22\pm 0.01$ \\
SAX/PDS       & this paper & 15--50 & CUTOFFPL$^e$ & $0.167\pm 0.017$& $1.4\pm 0.3$&
[41.13] & $5.88\pm 0.19$ & $9.22\pm 0.01$ \\
SAX/PDS       & this paper & 15--50 & CUTOFFPL$^e$ & $0.148\pm 0.002$& [1.29]      &
[41.13] & $5.43\pm 0.17$ & $9.22\pm 0.01$ \\
\enddata

\tablerefs{(1) \citet{Horstman75}; (2) \citet{Kinzer78}; (3) M80; (4) G99}

\tablenotetext{a}{$I(E) = N
E^{-\Gamma}$~photons~~cm$^{-2}$~s$^{-1}$~keV$^{-1}$~sr$^{-1}$}
\tablenotetext{b}{Bremsstrahlung as described in the XSPEC user manual with $N = K$
and $kT$ is the plasma temperature.}
\tablenotetext{c}{$I(E) = N E^{-\Gamma}
\exp{(-E/E_c)}$~photons~~cm$^{-2}$~s$^{-1}$~keV$^{-1}$~sr$^{-1}$}
\tablenotetext{d}{$I(E) = N
(E/20)^{-\Gamma}$~photons~~cm$^{-2}$~s$^{-1}$~keV$^{-1}$~sr$^{-1}$}
\tablenotetext{e}{$I(E) = N (E/20)^{-\Gamma}
\exp{(-E/E_c)}$~photons~~cm$^{-2}$~s$^{-1}$~keV$^{-1}$~sr$^{-1}$}
\tablenotetext{f}{In units of $10^{-8}$ erg cm$^{-2}$ s$^{-1}$ sr$^{-1}$}
\tablenotetext{g}{In units of $10^{-9}$ erg cm$^{-2}$ s$^{-1}$}

\end{deluxetable}

\clearpage


\newpage

\newpage

\newpage

\newpage

\newpage



\newpage

\newpage

\end{document}